\documentclass[aps,prl,twocolumn,a4paper,floatfix,showpacs]{revtex4-1}
\usepackage{amsmath,amssymb}
\usepackage{graphicx}
\usepackage{dcolumn}
\usepackage{bm}
\usepackage{subfigure}
\usepackage[dvips]{color}

\begin{document}

\title{Exchange Bias 
driven by Dzyaloshinskii-Moriya interactions}

\author{R. Yanes$^{1}$, J. Jackson$^{2}$, L. Udvardi$^{3}$,  L. Szunyogh$^{3}$ and U. Nowak$^{1}$}
\affiliation{$^1$Department of Physics, University of Konstanz, Germany \\
             $^2$Max-Planck Institute for Solid State Research, Stuttgart, Germany \\
             $^3$Department of Theoretical Physics and Condensed Matter Research Group of the Hungarian Academy of Sciences, Budapest University of Technology and Economics, Budapest, Hungary
}

\begin{abstract}
The exchange bias effect in compensated IrMn$_{3}$/Co($111$) system is studied using multiscale modelling from \textit{ab-initio} to atomistic calculations. We evaluate numerically the out-of-plane hysteresis loops of the bilayer for different thicknesses of the ferromagnet layer. The results show the existence of a perpendicular exchange bias field and an enhancement of the coercivity of the system. 
In order to elucidate the possible origin of the exchange bias, we analyse the hysteresis loops of a selected bilayer by tuning the different contributions to the exchange interaction across the interface. Our results indicate that the exchange bias is primarily induced by Dzyaloshinskii-Moriya interactions, while the coercivity is increased mainly due to a \textit{spin-flop} mechanism.
\end{abstract}


\maketitle

\textit{Introduction --}
The exchange bias effect is commonly found in magnetic heterostructures where a ferromagnet (FM) or ferrimagnet (FI) is in contact with an antiferromagnet (AFM). In such systems the exchange interaction between the FM (FI) and the AFM may induce a unidirectional anisotropy in the ferromagnet, which is reflected in the hysteresis loops by a shift along the magnetic field axis. This effect is called exchange bias.\cite{meiklejohnPR56} Over more than 50 years several theories have been developed to explain the exchange bias effect.\cite{reviewsEB} Most of these theories assume uncompensated spins at the interface of the antiferromagnet to pin the ferromagnet and, therefore, fail to explain the origin of the exchange bias in a system with a totally compensated interface. In order to cure this problem exchange bias (EB) models based on a domain state in diluted antiferromagnets due to an
imbalance of the number of impurities,\cite{nowakPRB02,miltenyiPRL00}  spin-flop coupling, \cite{koonPRL97} biquadratic exchange interaction, \cite{dimitrovPRB98} formation of domain walls,\cite{suessPRB03} or anisotropic exchange interactions across the interface\cite{ledermanPRB04} were developed.  In the last few years, considering only symmetry properties, the Dzyaloshinskii-Moriya (DM) interactions have been proposed as possible mechanism responsible to the exchange bias in compensated systems, e.g. for Fe$_{3}$O$_{4}$/CoO  multilayers \cite{ijiriPRL07} and in biasing systems with G-Type antiferromagnetic perovskite. \cite{dongPRL09}

The L1$_2$-type IrMn$_{3}$ is a triangular AFM which presents a non-collinear spin ground state, called T1 N\'eel state. It exhibits a large second-order magnetic anisotropy due to anisotropy in the exchange interaction. This high effective anisotropy entails an 
easy plane ($111$), to which the ground state is confined. \cite{laszloPRB09} When  IrMn$_{3}$ is capped by fcc Co, the magnetic moments, the magnetic anisotropy and the exchange interactions of both the AFM and FM are modified close to the interface. These effects are described in details in Ref.~\onlinecite{laszloPRB11}. In particular, sizeable DM interactions arise between the Co
and Mn atoms owing to the breaking of inversion symmetry at the interface. The ($111$) interface is perfectly compensated, i.e., there are equal numbers of atoms belonging to the three magnetic sublattices in the AFM. These properties make the IrMn$_{3}$/Co($111$) system a perfect model to study the exchange bias in compensated interfaces.

In this letter we focus on addressing the origin of the exchange bias in compensated IrMn$_{3}$/Co($111$) bilayers by performing numerical calculations of the hysteresis loops and selecting the roles played by different types of exchange interactions between Mn and Co atoms in the exchange bias. We find that the IrMn$_{3}$/Co($111$) displays perpendicular exchange bias effect. The main mechanism responsible to the  perpendicular EB 
in this system is the DM interactions, nevertheless, there are other minor contributions to the exchange bias due to the anisotropy in the exchange interactions through the interface. 

\textit{Model --}
We suppose that the magnetic properties of the bilayer system can be described within a
generalized Heisenberg model,
\begin{equation}\label{eq:Ham}
\mathcal{H}=-\frac{1}{2}\sum_{i,j}\vec{s}_{i}\boldsymbol{J}_{ij}\vec{s}_{j}
-\sum_{i}\vec{s}_{i}\boldsymbol{K}_{i}\vec{s}_{i}-\sum_{i}\mu_{i}\vec{H}_{A}\vec{s}_{i} \; ,
\end{equation}
where $\vec{s}_{i}$ represents a classical spin, i.e. a unit vector along the direction of the magnetic moment at site $i$, occupied by  either Cobalt or Manganese atoms. The first term stands for the exchange contribution to the energy, with $\boldsymbol{J}_{ij}$ denoting the tensorial exchange interaction. The second term comprises the one-site anisotropy and the magnetostatic energy, and $\boldsymbol{K}_{i}$ is called the anisotropy matrix.  In the presence of an external magnetic field, $\vec{H}_{A}$,
the last term adds a Zeeman contribution to the Hamiltonian, where $\mu_{i}$ is the magnetic moment of the atom $i$.

 The exchange interaction $\boldsymbol{J}_{ij}$  can further be decomposed into three terms,\cite{udvardiPRB03}
\begin{equation}\label{eq:Jij}
\boldsymbol{J}_{ij}=J_{ij}^{iso}\boldsymbol{I}+\boldsymbol{J}_{ij}^{S}+\boldsymbol{J}_{ij}^{A} \; ,
\end{equation}
 with $J_{ij}^{iso}=\frac{1}{3} Tr\Bigl[\boldsymbol{J}_{ij}\Bigr]$ the isotropic exchange interaction,  $\boldsymbol{J}_{ij}^{S}=\frac{1}{2}(\boldsymbol{J}_{ij}+\boldsymbol{J}_{ij}^{T})-J_{ij}^{iso}\boldsymbol{I}$ the traceless symmetric (anisotropic) part  and $\boldsymbol{J}_{ij}^{A}=\frac{1}{2}(\boldsymbol{J}_{ij}-\boldsymbol{J}_{ij}^{T})$ the antisymmetric part of the exchange tensor.
The latter one is clearly related to the Dzyaloshinskii-Moriya (DM) interaction,
 \begin{equation}
\vec{s}_{i}\boldsymbol{J_{ij}^{A}}\vec{s}_{j}=\vec{D}_{ij}(\vec{s}_{i}\times\vec{s}_{j}) \; ,
 \end{equation}
where $\vec{D}_{ij}$ is termed the DM vector.  The DM interaction arises due to the spin-orbit coupling and favours a perpendicular alignment of the spins $\vec{s}_{i}$ and $\vec{s}_{j}$. \cite{dzyaloshinskiiJPCS58, moriyaPR60} It vanishes if the system is centrosymmetric, however, for solids with complex lattices or  at interfaces and surfaces, where the inversion symmetry is broken, the DM interaction might play an important role. \cite{crepieuxJMMM98}

\begin{figure}
 \centering
\subfigure[]{\label{fig:sketch1}\includegraphics[width=0.175\textwidth]{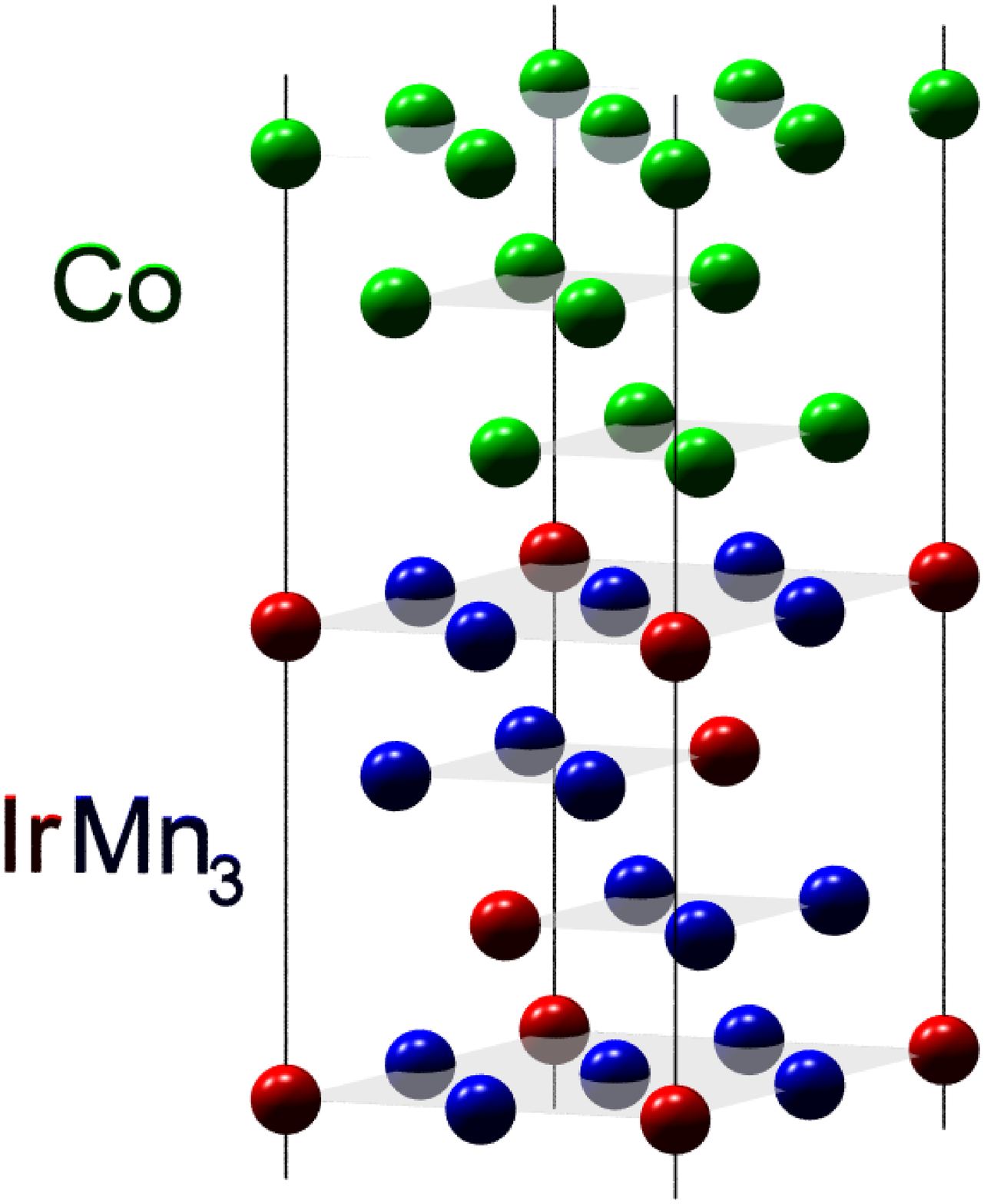}}
 \subfigure[]{\label{fig:geo1}\includegraphics[width=0.3\textwidth]{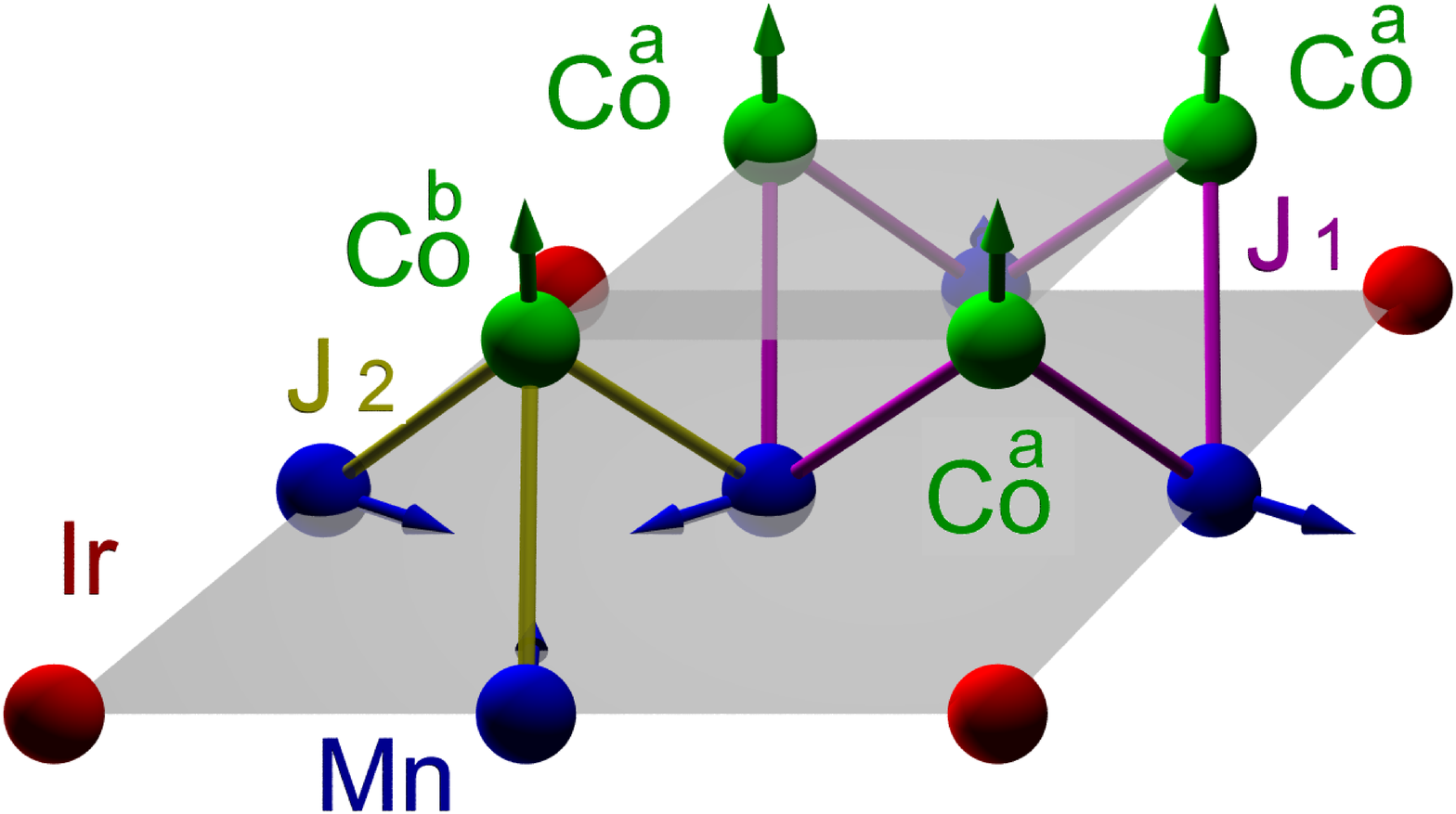}}
 \subfigure[]{\label{fig:geo2}\includegraphics[width=0.4\textwidth]{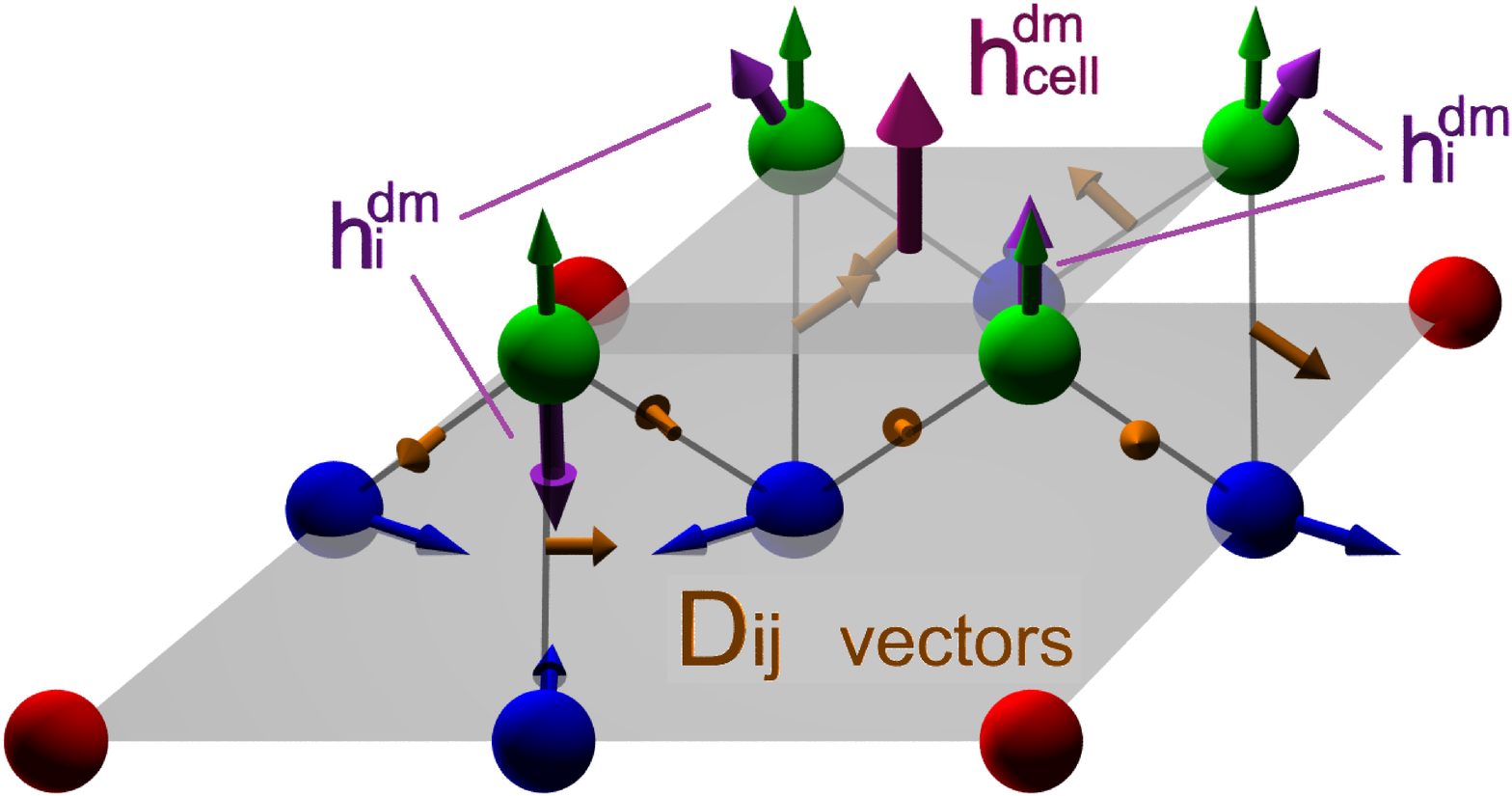}}
 \caption{(Color online) (a) Stacking order for the L1$_{2}$ IrMn$_{3}$ -- fcc Co bilayer near the interface. Red, blue and green spheres represent the Ir, Mn and Co atoms, respectively.  (b) Sketch of the magnetic order at the $(111)$ interface between IrMn$_{3}$ and Co. The moments of the Mn atoms (blue arrows) display a perfect T$1$ state. The AFM ($J_{1}$) and FM ($J_{2}$) character of the exchange interaction between the Co and Mn neighbours are also indicated. (c) The Dzyaloshinskii-Moriya vectors $\vec{D}_{ij}$ vectors  between Co-Mn nearest neighbours (orange arrows). The DM interface field, $\vec{h}_{i}^{\rm{dm}}$, and the DM field acting on
  the Co atoms per unit cell, $\vec{h}^{\rm{dm}}_{\rm{cell}}$, are displayed by violet and magenta arrows, respectively.} \label{fig:Geometria}
\end{figure}

 In terms of the fully relativistic Screened Korringa-Kohn-Rostoker (SKKR) method,\cite{laszloPRB94, zellerPRB95} we performed self-consistent calculations of a bilayer IrMn$_{3}$/Co($111$)  and, based on a spin-cluster expansion (SCE) technique, we derived the exchange interactions at the AFM/FM interface.\cite{laszloPRB11}
 For simplicity, we assumed that the fcc Co and the L1$_{2}$ IrMn$_{3}$ lattices match perfectly, i.e. no structural relaxation was considered and we used the two-dimensional lattice parameter of IrMn$_{3}$, $a=0.3785$ nm.
 In Fig.~\ref{fig:sketch1}, the atomic structure in the stacking model we used for the L1$_{2}$ IrMn$_{3}$-Co(fcc) bilayer is depicted.

As an insignificant simplification, for the spin-dynamics simulations we supposed that the Ir atoms are nonmagnetic, and the atomic magnetic moments of
 Mn and Co atoms were taken uniformly $\mu_{\rm{Mn}}=2.2\mu_{\rm{B}}$ and $\mu_{\rm{Co}}=1.6\mu_{\rm{B}}$.
 Due to symmetry, the three Mn sublattices in L1$_{2}$ IrMn$_{3}$ present uniaxial on-site anisotropy
 with different easy axes,\cite{laszloPRB09} however, with the same uniaxial magnetic anisotropy constant, 
$K_{\rm{Mn}}=0.54$ meV, as obtained from our first principles calculations. 
 The fcc Co has an one-site anisotropy constant smaller than $1 \, \mu$eV, therefore, in our simulations we neglected this contribution. Nevertheless, we considered magnetostatic interaction in the FM layer approximated by a uniaxial shape anisotropy  $K_{\rm{Co}}=-0.084$ meV. 

 \textit{Interface interactions --}
   As is well known, exchange bias is intimately related to the exchange interaction across the interface between the FM and the AFM part of the system. Unfortunately, it is in practice impossible to explore these interactions experimentally. First principles calculations represent, however, a powerful tool to determine the exchange interactions between magnetic atoms.

 In Fig.~\ref{fig:geo1}, the 2D unit cell is sketched at the interface, comprising four Co atoms in the upper plane, while three Mn and one Ir atoms in the below layer. Apparently, we can distinguish between two kinds of Co atoms at the interface: Co atoms with two or three nearest neighbour (NN) Mn atoms, Co$^{\rm{a}}$ and Co$^{\rm{b}}$, respectively. The Co$^{\rm{a}}$ atoms and their Mn NN's interact ferromagnetically, with an isotropic exchange constant $J_{1}=1.24$ meV. On the contrary, Co$^{\rm{b}}$  and their Mn NN's display an antiferromagnetic exchange interaction, $J_{2}=-6.37\,$meV. Therefore, considering even farther neighbour interactions,  
 the effective isotropic exchange interaction across the interface has an antiferromagnetic character.

 From Eq.~(\ref{eq:Ham}), the contribution to the interfacial energy due to the DM interactions, $\mathcal{H}^{\rm{DM}}_{\rm{int}}$, can be expressed as
\begin{equation}
\mathcal{H}^{\rm{DM}}_{\rm{int}}=\sum_{i,j}\vec{s}_{i}\cdot(\vec{s}_{j} \times \vec{D}_{ij})=\sum_{i}\vec{s}_{i}\cdot\vec{h}^{\rm{dm}}_{i}, \;
\end{equation}
where now the index $i$ labels only Co atoms and $j$ labels their Mn nearest neighbours. Correspondingly, $\vec{h}^{\rm {dm}}_{i}$ is the DM field experienced by a Co spin $\vec{s}_{i}$ due to its Mn neighbours.
As shown in Fig.~\ref{fig:geo2}, the DM vectors between the Co atoms and their Mn NN's lie practically in the ($111$) plane, with the magnitudes $|\vec{D}_{ij}|=0.58\,$meV and $|\vec{D}_{ij}|=0.42\,$meV for Co$^{\rm{a}}$-Mn and the Co$^{\rm{b}}$-Mn nearest neighbours, respectively.

To simplify the discussion of the exchange bias effect, let us consider only nearest neighbour interactions and suppose that the Co spins are in a ferromagnetic state and the Mn atoms form a perfect T$1$ state.  
Then the DM interactions across the interface induce an effective magnetic field acting on the Co atoms per unit cell, $|\vec{h}^{\rm{dm}}_{\rm{cell}}|=|\sum_{i=1}^{4} \vec{h}^{\rm{dm}}_{i}|\approx 1.05$ meV, 
that points normal to the interface. Concomitantly, the DM interactions across the interface favour a perpendicular alignment of the Co and Mn moments. The direction of $\vec{h}^{\rm{dm}}_{\rm{cell}}$, pointing either towards the
  Co or the IrMn$_3$ part of the interface, depends on the chirality of the T$1$ state.
Supposing that during the magnetic reversal there is no distortion in the magnetic state of the Co and the Mn spins, 
an exchange bias arises due to the DM interactions between the Co and Mn neighbours with the EB field,
\begin{equation}\label{eq:Heb}
H_{\rm{eb}}^{\rm{dm}}=\frac{|\vec{h}^{\rm{dm}}_{\rm{cell}}|\cos{\theta}}{4\mu_{\rm{Co}}t_{\rm{Co}}} \: ,
\end{equation}
where $\theta$ is the angle between the direction of the magnetization in the FM and the effective DM field,
and $t_{\rm{Co}}$ is the thickness of the Co layer.

\textit{Results and discussions --} In the spin-dynamics simulations the antiferromagnet was modelled by three intercalated Mn sublattices, forming in total $20\times20\times6$ unit cells, and the ferromagnet with $20\times20\times t_{\rm{Co}}$ unit cells, $t_{\rm{Co}}$ denoting the number of Co atomic monolayers. As what follows the bilayers will be labelled  by [IrMn$_{3}$]$_{6}$/[Co]$_{t_{\rm{Co}}}$.

In order to study the possible existence of the exchange bias effect we evaluate numerically the hysteresis loops of several [IrMn$_{3}$]$_{6}$/[Co]$_{t_{\rm{Co}}}$ bilayers. The hysteresis loops are simulated by solving the Landau-Lifshitz-Gilbert (LLG) equation in context of the generalized anisotropic Heisenberg model described in Eq.~(\ref{eq:Ham}), where the exchange interactions are considered up to sixth NN's. In these simulations the applied field pointed parallel to [$111$] direction, i. e., perpendicular to the interface (along the $z$ axis).

Prior to calculating the hysteresis loops, we prepared the system in a similar way as in the experiment, namely, by simulating a field-cooling (FC) process. The FC process starts from a random spin configuration in the AFM part, at an initial temperature T above the N\'{e}el temperature of the AFM and below the Curie temperature of the FM, and proceeds to a final temperature  T$_{f}=0$~K. This process is done under the influence  of an external applied (cooling) field, $H_{cf}$=1.5~T.
After the FC process the magnetic moments in the FM are oriented along the direction of the cooling field and perpendicular to the AFM easy plane. The AFM presents a quasi-T1 state, slightly distorted at the interface due to the effective antiferromagnetic interaction between the Co and Mn atoms. This distortion gives rise to a small net magnetization in the AFM that is anti-parallel to the FM magnetization. Therefore, the spin configuration after the simulated FC process is similar to a \textit{spin-flop} state.

  \begin{figure}
    \centering
      \includegraphics[width=0.3\textwidth]{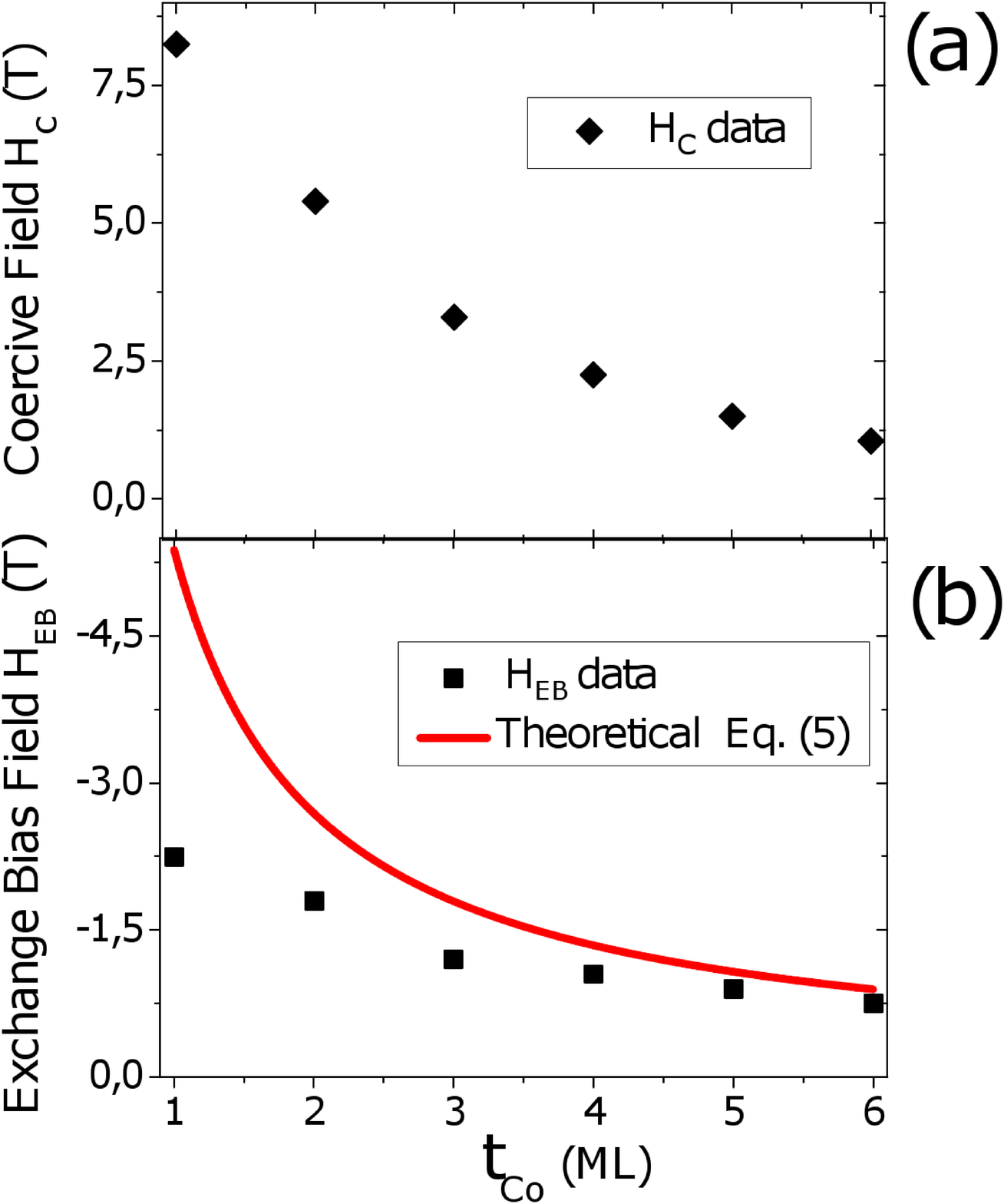}
     \caption{(Color online) Dependence of the (a) coercive field  and (b) exchange bias with the thickness of Co capping $t_{\rm{Co}}$ in ML. The numerical results are presented by the solid symbols and the solid line displays the theoretical decay function of the exchange bias field,  Eq.~(\ref{eq:Heb}).} \label{fig:Hc_Heb_tco}
   \end{figure}
 
 Using this spin configuration as the initial magnetic state, the simulated hysteresis loops display a quasi-square shape, negative exchange bias (H$_{{EB}}$) and a high coercivity, for a Cobalt thickness range from 1 to 6 atomic monolayers.
 In Fig.~\ref{fig:Hc_Heb_tco}(a) we show the values of the coercive field determined from the out-of-plane hysteresis loops as a function of $t_{\rm{Co}}$. The system displays a high coercivity which decays as $t_{\rm{Co}}$ increases and fit relatively well a $1/t_{\rm{Co}}$ dependence. This result clearly indicates that the large coercive field  is attributed to the interface.  
 
 In Fig.~\ref{fig:Hc_Heb_tco}(b) we present the values of the exchange bias field  for several values of the Co thickness $t_{\rm{Co}}=1-6$ ML. $H_{\rm{EB}}$ is numerically determined as the shift of the center of the hysteresis loops. As expected the exchange bias field decreases as the Co thickness is increases, but the fit of our calculated data to the theoretical expression, Eq.~(\ref{eq:Heb}), is not perfect, see the red solid line in  Fig.~\ref{fig:Hc_Heb_tco}(b). This deviation from the theoretical $1/t_{\rm{Co}}$ decay can be understood by noting that the influence of the interface is not restricted to the adjacent FM and AFM atomic layers, but it extends to at least two-three atomic layers from the interface on both sides.

Based on the spin-dynamics simulations we can affirm that the FM magnetization switches through a quasi-coherent rotation, and during the FM switching the AFM also switches between two \textit{spin-flop} states. As it was shown by Schulthess and Butler in Ref.~\onlinecite{schulthessPRL98}, if only isotropic exchange interactions are considered, this kind of magnetization inversion process leads to a uniaxial rather than unidirectional anisotropy. Therefore, the high value of the coercive field might be explained with a \textit{spin-flop} mechanism, but it is not related to the exchange bias effect.

Inspecting the effect of the DM interactions during the magnetization reversal, we note that the out-of-plane component of the interfacial DM field only depends on the in-plane components of the AFM magnetization which practically remain unchanged during the switching process. This implies that in the descending branch of the hysteresis loop  $\vec{h}_{\rm{cell}}^{\rm{dm}}$ opposes the switching of the FM and in the ascending branch it favours the inversion of the FM magnetization. This is the simple picture how the interfacial DM field generates the perpendicular exchange bias in the IrMn$_{3}$/Co bilayer.

\begin{figure}
 \includegraphics[width=9cm]{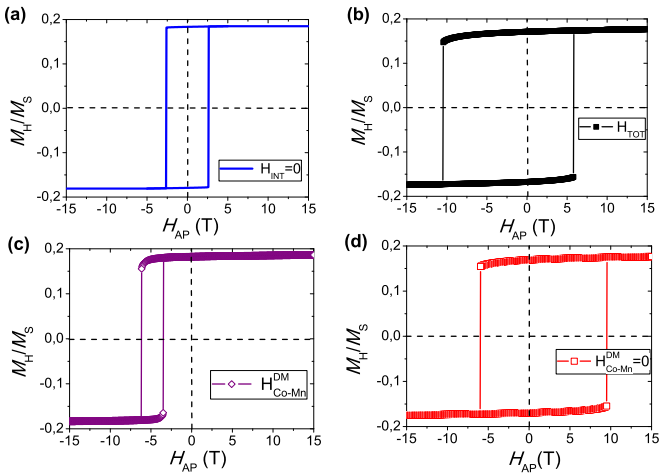}
 \caption{(Color online) Hysteresis out-of-plane loops of [IrMn$_{3}$]$_{6}$/[Co]$_{1}$: (a) Removing all the interactions across the interface. (b)  Considering all the interactions across the interface. (c) Considering only the DM interactions between the Co and Mn atoms. (d) Removing only the DM interaction between the Co and Mn atoms.}\label{fig:Hyst_Jcont}
\end{figure}

In order to confirm our hypothesis for the origin of the exchange bias, in case of the  bilayer [IrMn$_{3}$ ]$_{6}$/[Co]$_{1}$ we investigated the change in the hysteresis loop when switching on or off the different contributions to the exchange interactions, Eq.~(\ref{eq:Jij}), between FM and AFM moments. In all  cases the initial magnetic state of the simulation was the same, namely, the spin configuration obtained after a FC process with all the interactions between FM and AFM layers switched on.

If all parts of the exchange interactions is removed between Co and Mn atoms, the hysteresis loop presents a perfect square shape and, as expected, does not display any EB shift, see Fig. \ref{fig:Hyst_Jcont}(a). In this case the coercive field is $H_{\rm{C}}=2.65\,$T.  This value of $H_{\rm{C}}$ is larger than the expected $H_{\rm{C}}$ value considering only the shape anisotropy, $K_{\rm{Co}}$. It is also interesting that the shape of the hysteresis loops indicates that the system displays a perpendicular anisotropy. Such perpendicular anisotropy and large coercivity may be explained by an increase of the symmetric anisotropic part of the exchange interactions between the Co atoms close to the interface.\cite{laszloPRB11}

Fig.~\ref{fig:Hyst_Jcont}(b) shows the hysteresis loop of the selected bilayer when taking into account all the contributions to the Hamiltonian (\ref{eq:Ham}). Apparently, the system exhibits a highly enhanced coercive field ($H_{\rm{C}}=8.2\,$T) and high negative exchange bias ($H_{\rm{EB}}=-2.3\,$T). In comparison to the previous case, it is obvious that both features are related to the exchange interactions across the FM/AFM interface.

Considering only the asymmetric DM contributions to the exchange interactions across the interface, the hysteresis loop of the system shows a drastic reduction of the coercive field and an increase of the exchange bias with respect to the case when all the interactions are included, see Fig.~\ref{fig:Hyst_Jcont}(c). Since in this case the isotropic exchange interaction between the Co and Mn atoms is switch off, there is no distortion of the AFM T1 state close to the interface. Consequently, the net AFM magnetization becomes zero close to the interface, preventing thus the increment of the coercive field.

It is interesting to compare the value of the exchange bias obtained in the simulation, $H_{\rm{EB}}=-4.07\,$T, with the effective exchange bias field given by the Eq.~(\ref{eq:Heb}) for $t_{\rm{Co}}=1\,$~ML, $H_{\rm{eb}}^{\rm {dm}}\approx-5.38\,$T. These values are of the same order and sign, the model value is, however, considerably larger in magnitude than $H_{\rm{EB}}$. This difference in the theoretical and numerical values may be a consequence of neglecting the DM interactions between Co-Co and Mn-Mn neighbours in the theoretical model.    

Fig.~\ref{fig:Hyst_Jcont}(d) presents the calculated hysteresis loop of the given bilayer when only the DM interactions between the Mn and Co atoms are removed. In this case, the coercivity is almost not affected ($H_{\rm{C}}\approx 7.8\,$~T). Nevertheless, a strong reduction in the magnitude of the exchange bias is observed and even its sign is changed ($H_{\rm {EB}}\approx 1.8\,$~T). This suggests that beyond the DM interactions between the Co and Mn moments there are other, less significant, sources of the exchange bias. These mechanisms can be related to the DM interactions between the Mn spins close to the interface and also the anisotropic exchange interactions across the interface. 

In comparison to experimental values of perpendicular exchange bias field (e.g. in IrMn/[Co/Pt]\cite{sortPRB05,liuJPDAP09} and IrMn/Co/[Co/Pt]\cite{wuJAP13} multilayers) our results are 20-100 times higher, but we should consider that these experimental data are evaluated at room temperature, while ours simulation of perpendicular EB are calculated at 0 K. Also it is possible that the Mn diffusion deteriorates the first Co atomic layer,\cite{liuJPDAP09} and the chemical disorder decreases the DM interactions between Mn and Co atoms at the interface leading to a reduction of the EB. On the other hand, in the literature values have been reported of in-plane exchange bias field in Mn-Ir/Co$_{100-x}$Fe$_{x}$ bilayers \cite{imakitaAPL04,tsunodaAPL10} of the same order of magnitude as our prediction for a Co layer with a thickness of 4 nm.

\textit{Conclusion --}
In the light of our results in terms of combined \textit{ab initio} and spin-dynamics simulations we can conclude that the principal source of the perpendicular exchange bias of the IrMn$_{3}$/Co($111$) is the Dzyaloshinskii-Moriya interaction across the interface that favours a perpendicular orientation between the Mn and Co moments and a unidirectional anisotropy perpendicular to the interface.  The high coercive field we obtained from our simulations is due to a combination of at least two factors. One reason is the enhanced two-site anisotropy (symmetric anisotropic part of the exchange) of the Co atoms close to the interface. More importantly, however, the isotropic exchange between the Mn and Co atoms results in a distortion of the T1 spin structure close to the interface, which leads to a net magnetization in the AFM interface layer. During the switching of the FM, the AFM also switches between two \textit{spin-flop} states, resulting thus in a considerable enhancement of the coercivity.

Financial support was in part provided by the New Sz\'echenyi Plan of Hungary (Project ID. T\'AMOP-4.2.2.B-10/1-2010-0009) 
and by the Hungarian Scientific Research Fund  (contracts OTKA No. K77771 and K84078).  U.N. and
L.S. acknowledge financial support from the DFG through SFB 767.


\begin{thebibliography}{widest-label}
\bibitem{meiklejohnPR56} W. H. Meiklejohn and C. P. Bean, Phys. Rev. \textbf{102}, 1413 (1956).
\bibitem{reviewsEB} J. Nogu\'{e}s and I. K. Schuller, J. Magn. Magn. Mater. \textbf{192}, 203 (1999);  A. Berkowitz, and K.Takano, J. Magn. Magn. Mater. \textbf{200}, 552 (1999); R. L. Stamps, J. Phys. D: Appl. Phys. \textbf{33}, R247-R268 (2000); M. Kiwi, J. Magn. Magn. Mater. \textbf{234}, 584 (2001).
\bibitem{nowakPRB02} U Nowak, K. D. Usadel, J. Keller, P. Milt\'{e}nyi, B. Beschoten, and G. G\"untherodt, Phys. Rev. B \textbf{66}, 014430 (2002).
\bibitem{miltenyiPRL00} P. Milt\'{e}nyi, M. Gierlings, J. Keller, B. Beschoten, G. G\"{u}ntherodt, U. Nowak, and K. D. Usadel, Phys. Rev. Lett. \textbf{84}, 4224 (2000).
\bibitem{koonPRL97}  N. C. Koon, Phys. Rev. Lett. \textbf{78}, 4865 (1997).
\bibitem{dimitrovPRB98} D. V. Dimitrov, S. Zhang, J. Q. Xiao, G. C. Hadjipanayis, and C. Prados, Phys. Rev. B \textbf{58}, 12090 (1998).
\bibitem{suessPRB03} D. Suess, M.Kirschner, T. Schrefl, J. Fidler, R. L. Stamps, and J.V. Kim, Phys. Rev. B \textbf{67}, 054419 (2003).
\bibitem{ledermanPRB04} D. Lederman, R. Ram\'{i}rez, and M. Kiwi, Phys. Rev. B \textbf{70}, 184422 (2004).
\bibitem{ijiriPRL07} Y. Ijiri, T. C. Schulthess, J. A. Borchers, P. J. van der Zaag and R. W. Erwin, Phys. Rev. Lett. \textbf{99}, 147201 (2007).
\bibitem{dongPRL09} S. Dong, K. Yamauchi, S. Yunoki, R. Yu, S. Liang, A. Moreo, J-M Liu, S. Picozzi, and E. Dagotto, Phys. Rev. Lett. \textbf{103}, 127201 (2009).
\bibitem{laszloPRB09} L. Szunyogh, B. Lazarovits, L. Udvardi, J. Jackson, and U. Nowak, Phys. Rev. B \textbf{79} 020403(R) (2009).
\bibitem{laszloPRB11} L. Szunyogh, L. Udvardi, J. Jackson, U. Nowak, and R. Chantrell, Phys. Rev. B \textbf{83} 024401 (2011).
\bibitem{dzyaloshinskiiJPCS58} I. Dzyaloshinskii, J. Phys. Chem. Solids \textbf{4}, 241 (1958).
\bibitem{moriyaPR60} T. Moriya, Phys. Rev. \textbf{120}, 91 (1960).
\bibitem{crepieuxJMMM98} A. Cr\'{e}pieux, and C. Lacroix, J Magn. Magn Mater. \textbf{182}, 341 (1998).
\bibitem{laszloPRB94} L. Szunyogh, B Ujfalussy, P. Weinberger, and J. Koll\'ar, Phys. Rev. B  \textbf{49}, 2721 (1994).
\bibitem{zellerPRB95} R. Zeller, P. H. Dederichs, B. Ujfalussy, L. Szunyogh, and P. Weinberger, Phys. Rev. B \textbf{52}, 8807 (1995).
\bibitem{udvardiPRB03} L. Udvardi, L. Szunyogh, K. Palot\'as, and P. Weinberger, Phys. Rev. B \textbf{68}, 104436 (2003).
\bibitem{schulthessPRL98}  T. C. Schulthess and W. H. Butler, Phys. Rev. Lett. \textbf{81}, 4516 (1998).
\bibitem{sortPRB05} J. Sort, V. Baltz, F. Garcia, B. Rodmacq, and B. Dieny, Phys. Rev. B \textbf{71}, 054411 (2005).
\bibitem{liuJPDAP09} Y. F. Liu, J. W. Cai, and S. L. He, J. Phys. D: Appl. Phys. \textbf{42}, 115002 (2009).
\bibitem{wuJAP13} Qiong Wu, Wei He, Hao-Liang Liu, Yi-fan Liu, Jian-Wang Cai, and Zhao-Hua Cheng, J. Appl. Phys. \textbf{113}, 033901 (2013).
\bibitem{imakitaAPL04}  K. Imakita, M. Tsunoda, and M. Takahashi, Appl. Phys. Lett. \textbf{85}, 3812 (2004).  
\bibitem{tsunodaAPL10} M. Tsunoda, H. Takahashi, T. Nakamura, C. Mitsumata, S. Isogami, and M. Takahashi, Appl. Phys. Lett. \textbf{97}, 072501 (2010).   
\end{thebibliography}
\end{document}